\RequirePackage{fixltx2e}
\documentclass[a4paper]{isp}
\usepackage[american]{babel}
\usepackage{graphicx}
\usepackage{paralist}
\usepackage[T1]{fontenc}
\usepackage{lmodern}
\usepackage{microtype}
\usepackage{amsmath}
\usepackage{booktabs}
\usepackage{url}
\ifnum\pdfoutput>0
\usepackage[
bookmarks=false,
breaklinks=true,
colorlinks=true,
linkcolor=black,
citecolor=black,
urlcolor=black,
pdfpagelayout=SinglePage
]{hyperref}
\usepackage[all]{hypcap}
\else
\usepackage{hyperref}
\fi

\usepackage[capitalise,nameinlink]{cleveref}
\crefname{section}{Sect.}{Sect.}
\Crefname{section}{Section}{Sections}
\crefname{figure}{Figure}{Figure}
\Crefname{figure}{Figure}{Figures}

\usepackage{xspace}
\begin{document}
\pdfgentounicode=1

\title{Robust Video Watermarking\\against H.264 and H.265 Compression Attacks}
%If Title is too long, use \titlerunning
%\titlerunning{Short Title}

%Single insitute
%Currently disabled
%\author{Nematollah~Zarmehi
%	and~Mohammad Javad~Barikbin% <-this % stops a space
%	\thanks{N. Zarmehi is with the Department
%		of Electrical Engineering, Sharif University of Technology, Tehran,
%		Iran, (e-mail: zarmehi\_n@sharif.edu).}% <-this % stops a space
%	\thanks{M. J. Barikbin is with the Department
%		of Electrical and Computer Engineering, University of Tehran, Tehran,
%		Iran, (e-mail: mj.barikbin@ut.ac.ir).}% <-this % stops a space
%}

%\iffalse
%%\author{Nematollah Zarmehi \and  \and Rajesh Koothrappali \and Howard Wolowitz}
%%If there are too many authors, use \authorrunning
%%\authorrunning{First Author et al.}
%\institute{ Division of Physics, Mathematics and Astronomy, California Institute of Technology, Pasadena, United States}
%\fi

%Multiple insitutes
%Multiple institutes are typeset as follows:
\author{Nematollah Zarmehi\inst{1}\corauthor \and Mohammad Javad Barikbin\inst{2}}

%If there are too many authors, use \authorrunning
%\authorrunning{First Author et al.}

\institute{
Advanced Communication Research Institute, Electrical Engineering Department, Sharif University of Technology, Tehran, Iran\\
\and
Electrical and Computer Engineering Department, University of Tehran, Tehran, Iran\\
\email{http://zarmehi.ir/contact.html}
}
			
\maketitle

\begin{abstract}
This paper proposes a robust watermarking method for uncompressed video data against H.264/AVC and H.265/HEVC compression standards. We embed the watermark data in the mid-range transform coefficients of a block that is less similar to its corresponding block in the previous and next frames. This idea makes the watermark robust against the compression standards that use the inter prediction technique. The last two video compression standards also use inter prediction for motion compensation like previous video compression standards. Therefore, the proposed method is also well suited with these standards. Simulation results show the adequate robustness and transparency of our watermarking scheme.
\end{abstract}

\section{Introduction}
One of the main concerns of digital media producers is preventing copyright infringement. Nowadays the concerns in the digital media business are growing due to the advances in technology particularly digital communications \cite{bookw}. Prevention from copying or unauthorized publications is of vital importance to their manufacturers. For example it is possible that in a few seconds, large amounts of a digital media could be illegally published on the Internet \cite{zarmehi}. Watermarking is a solution for these issues. In watermarking by adding a watermark to the media it can be protected from illegal copying \cite{shih}. Also for authentication purposes we can prove the ownership of the media by extracting the hidden watermark in it. Digital video is one of the most important digital media today that many watermarking, steganography, and steganalysis methods are proposed for video \cite{srv1,srv2,srv3,srv4}. In many applications like military and cinema improving security of raw videos is of high importance and watermarking can be used to achieve this objective. The main problem of this process is the robustness of the watermark against the usual attacks. The most common attack against a raw video is compression. Robustness of the watermarking should be to the extent that if the watermarked video would have been compressed and subsequently decompressed, the watermark would still remain in the video.

In \cite{park}, the authors proposed a video watermarking method, using video characteristics based on Human Visual System (HVS) in 3D-DCT domain. They embedded the watermark in the mid-range 3D-DCT coefficients and evaluated the robustness of their method against MPEG-2 compression attack. Alavianmehr et al. \cite{alavani} proposed a reversible data hiding method for uncompressed video data based on histogram distribution constraint. 

Recently, various embedding methods for raw video are proposed. In \cite{masking}, a watermarking in 3D-DWT domain is proposed using perceptual masking. The watermark is embedded additively by weighing it through the predefined mask. 

In \cite{obvious}, the temporal sensitivity of the human visual system (HVS) is utilized to exploit the temporal dimension for watermark embedding. To achieve imperceptible distortion after watermark insertion, the embedding process uses temporal contrast thresholds of HVS to determine the maximum strength of watermark. For more details about the video watermarking methods in uncompressed domain please refer to \cite{srv}.

The latest standards for video compression are H.264/AVC \cite{h264} and H.265/HEVC \cite{h265} standards. In this paper, we propose a video watermarking method that is robust against compression by the aforementioned standards. To the extent of our knowledge, there has not been a method proposed for uncompressed video watermarking which is robust against H.265 compression at the moment.

\section{Proposed Method}
We embed the watermark in the DCT coefficients of some blocks of the luma frames. For this purpose, every block of a frame that is suitable for watermarking should be selected and after that the watermark data could be embedded in these selected blocks. We call each of the selected blocks a Watermarking Block (WB). In this section, first we introduce our method on how to choose the WBs and afterwards we present the embedding and extracting algorithms.

\subsection{Search for WBs}
To propose a method that is robust against H.264 and H.265 compression, we should know the basics of how they work that means by considering the compression process we can design a reliable watermarking method.
Inter prediction is used in both H.264 and H.265 standards. In the compression process, the encoder makes a prediction of the current block based on previously coded frames using inter prediction. Then, the prediction block is subtracted from the current block to form the residual block \cite{rechardson1}-\cite{rechardson2}. Therefore if a block and its predicted version are similar, the residual values will be small. To increase the robustness of our method, we choose those blocks of a frame that are less similar to their corresponding blocks in the previous and next frames.

Let ${\mathbf{B}_{i,k}}$ be the $i$-th block of the $k$-th frame. We calculate the residual blocks ${\mathbf{R}_{i,k}^{-1}}$ and ${\mathbf{R}_{i,k}^{+1}}$ as follows:

$$
{\mathbf{R}_{i,k}^{-1}} = {\mathbf{B}_{i,k}} - {\mathbf{B}_{i,k-1}}$$
\begin{equation}
{\mathbf{R}_{i,k}^{+1}} = {\mathbf{B}_{i,k}} - {\mathbf{B}_{i,k+1}}
\end{equation}
where ${\mathbf{R}_{i,k}^{-1}}$ and ${\mathbf{R}_{i,k}^{+1}}$ are the residual blocks regarding the corresponding block in the previous frame (${\mathbf{B}_{i,k-1}}$) and in the next frame (${\mathbf{B}_{i,k+1}}$), respectively. If the sum of energy of the two mentioned residual blocks is greater than a threshold $E_{th}$ then the block ${\mathbf{B}_{i,k}}$ will be a WB.

The drawback of the proposed method is its capacity. If the video has slow movements, we can find few WBs which lead to low capacity. To overcome this issue, choosing lower threshold $E_{th}$ is inevitable.

\begin{figure}[!t]
	\centering
	\includegraphics[width=1 \linewidth]{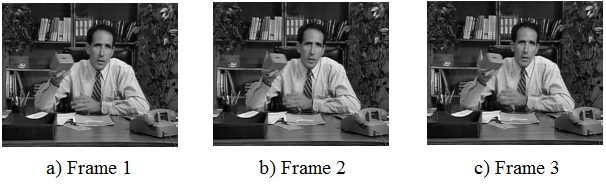}
	\caption{Three consequent frames of Salesman video sequence}
	\label{fig1}
\end{figure}

\begin{figure}[!t]
	\centering
	\includegraphics[width=1 \linewidth]{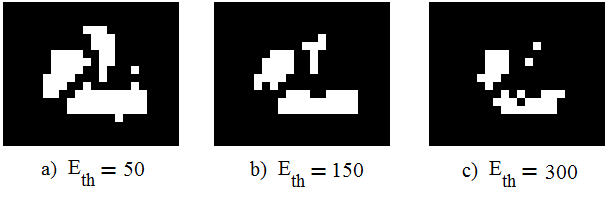}
	\caption{WBs of the second frame of Salesman video sequence for different values of $E_{th}$}
	\label{fig2}
\end{figure}

\begin{figure}[!t]
	\centering
	\includegraphics[width=1 \linewidth]{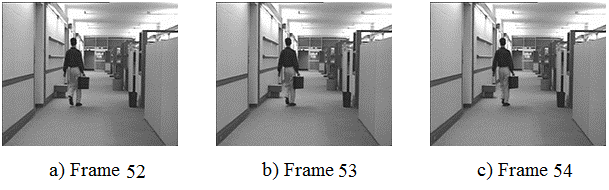}
	\caption{Three consequent frames of Salesman video sequence}
	\label{fig44}
\end{figure}

\begin{figure}[!t]
	\centering
	\includegraphics[width=1 \linewidth]{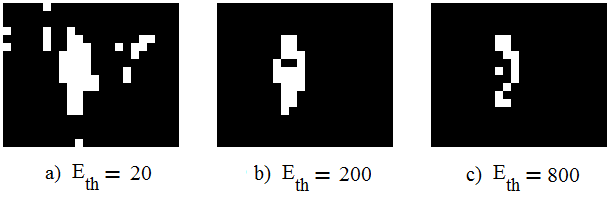}
	\caption{WBs of the 53rd frame of Hall video sequence for different values of $E_{th}$}
	\label{fig55}
\end{figure}

Figures \ref{fig1} and  \ref{fig44} show three consecutive frames of the Salesman and Hall video sequences. For different values of $E_{th}$, the WBs of the middle frame are displayed in white and the others are displayed in black in Figures \ref{fig2} and \ref{fig55}. In the frames shown in Figure \ref{fig1}, the head and hands of the salesman are moving and the proposed method correctly identifies these moving areas of the frame. It can be seen that the blocks that are similar in the consecutive frames are not selected and as $E_{th}$ increases, fewer blocks will be selected for watermarking. This claim is also true for the Hall video sequence.

An important advantage of this method is that by embedding in the moving areas of the frame, the watermark will be highly imperceptible.

\subsection{Watermark Embedding}
Let $X_{i,k}$ be the {\it i}-th WB of the k-th frame. First we calculate its DCT coefficients, then after zig-zag scanning the coefficients we pick two of the mid-range DCT coefficients $c_1$ and $c_2$ according to a pseudorandom key. The embedding rule is defined as:
\begin{equation}
\small{
	\left\{ {\begin{array}{*{20}{c}}
		{{\rm{For~embedding~bit~``1":~replace~}}\left| {{c_1}} \right|{\rm{ with }}\max \left\{ {\left| {{c_1}} \right|,\left| {{c_2}} \right|} \right\}}\\
		{{\rm{For~embedding~bit~``0":~replace~}}\left| {{c_1}} \right|{\rm{ with }}\min \left\{ {\left| {{c_1}} \right|,\left| {{c_2}} \right|} \right\}}
		\end{array}} \right.}
\end{equation}

\noindent This way one bit can be embedded in each WB.

\subsection{Watermark Extraction}
Considering the proposed embedding method, the extraction algorithm will be simple. First we locate the WBs of a luma frame. After calculating the DCT coefficients of the WBs and  zig-zag scanning, the watermarked coefficients $c_1^{\prime}$ and $c_2^{\prime}$ should be identified according to the pseudorandom key and then the watermark bit will be extracted as follows:
\begin{equation}
{\rm{If }}\left| {c_1^{\prime}} \right| > \left| {c_2^{\prime}} \right|{\rm{:~bit~``1",~Else:~bit~``0" }}{\rm{.}}
\end{equation}

\section{Simulation Results}
\begin{figure}[!t]
	\centering
	\includegraphics[width=0.25 \linewidth]{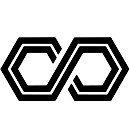}
	\vspace{0.2cm}
	\caption{The logo of Isaac Scientific Publishing (ISP) journals.}
	\label{fig3}
\end{figure}

For evaluating the performance of our watermarking system, we embedded the logo of Isaac Scientific Publishing (ISP) journals showed in Figure \ref{fig3}  into four video sequences and then compressed and decompressed the watermarked video with H.264/AVC and H.265/HEVC codecs. Tables \ref{tab1} and \ref{tab2} show the correlation between the original logo and the extracted one.

\begin{table}
	\renewcommand{\arraystretch}{1.3}
	\caption{Correlation between original watermark and the extracted one after compression and decompression with H.264/AVC}\label{tab1}
	\vspace{0.1cm}
	\centering
	\small{
		\begin{tabular}{|c|c|}
			\hline
			{\bf Video Sequence} & {\bf Correlation} \\ \hline
			Bus                             & 0.9901                       \\ \hline
			Table-tennis                    & 0.9827                       \\ \hline
			Foreman                         & 0.8256                       \\ \hline
			Football                        & 0.8469                       \\ \hline
		\end{tabular}}
	\end{table}
	
	\begin{table}[!h]
		\renewcommand{\arraystretch}{1.3}
		\caption{Correlation between original watermark and the extracted one after compression and decompression with H.265/HEVC}\label{tab2}
		\vspace{0.1cm}
		\centering
		\small{
			\begin{tabular}{|c|c|}
				\hline
				{\bf Video Sequence} & {\bf Correlation} \\ \hline
				Bus                             & 0.9677                       \\ \hline
				Table-tennis                    & 0.8628                       \\ \hline
				Foreman                         & 0.8049                       \\ \hline
				Football                        & 0.7258                       \\ \hline
			\end{tabular}}
		\end{table}

		We also verified the robustness of our watermarking scheme by more than one step compression and decompression. Tables \ref{tab3} and \ref{tab4} show the correlation between the original watermark and the extracted one after two consecutive compressions and decompressions by H.264/AVC and H.265/HEVC codecs.
		
		\begin{table}[!h]
			\renewcommand{\arraystretch}{1.3}
			\caption{Correlation between original watermark and the extracted one after two step compression-decompression with H.264/AVC}\label{tab3}
			\vspace{0.1cm}
			\centering
			\small{
				\begin{tabular}{|c|c|}
					\hline
					{\bf Video Sequence} & {\bf Correlation} \\ \hline
					Bus                             & 0.7559                       \\ \hline
					Table-tennis                    & 0.9542                       \\ \hline
					Foreman                         & 0.7254                       \\ \hline
					Football                        & 0.7526                       \\ \hline
				\end{tabular}}
			\end{table}

			\begin{table}[!h]
				\renewcommand{\arraystretch}{1.3}
				\caption{Correlation between original watermark and the extracted one after two step compression-decompression with H.265/HEVC}\label{tab4}
				\vspace{0.1cm}
				\centering
				\small{
					\begin{tabular}{|c|c|}
						\hline
						{\bf Video Sequence} & {\bf Correlation} \\ \hline
						Bus                             & 0.8693                       \\ \hline
						Table-tennis                    & 0.7670                       \\ \hline
						Foreman                         & 0.7845                       \\ \hline
						Football                        & 0.6118                       \\ \hline
					\end{tabular}}
				\end{table}
				
				The results of Tables \ref{tab3} and \ref{tab4} verify the robustness of the proposed watermarking method against compression and decompression attacks.
				
				The transparency of the proposed watermarking method is measured by PSNR. Park et al. \cite{park} proposed a video watermarking method using video characteristics based on HVS in 3D-DCT domain. We compared the transparency of our proposed method with the results of \cite{park} in TABLE \ref{tab5}. 
				
				\begin{table}[!h]
					\renewcommand{\arraystretch}{1.3}
					\caption{Correlation between original watermark and the extracted one after two step compression-decompression with H.265/HEVC}\label{tab5}
					\vspace{0.1cm}
					\centering
					\small{
						\begin{tabular}{|c|c|c|}
							\hline
							{\bf Video} & {\bf PSNR (dB)} & {\bf PSNR (dB)} \\ 
							{\bf Sequence} & {\bf (proposed method)} & {\bf \cite{park}} \\ \hline
							Table-tennis                    & 48.41                 & 42.42                 \\ \hline
							Foreman                         & 49.58                 & 44.48                 \\ \hline
							Football                        & 50.94                 & 42.32                 \\ \hline
						\end{tabular}}
					\end{table}
					
					%{\bf Interpolation} &\multicolumn{2}{ |c| }{{\bf PSNR (dB) at Different Sampling Rates}}  \\ \cline{2-3}
					%    {\bf Method} &  ~~~~~~${\boldsymbol f_s}$~~~~~~ &  ~~~~~~${\boldsymbol {2f_s}}$ ~~~~~~  \\ \hline
					
\begin{figure}[!t]
						\vspace{0.2cm}
						\centering
						\includegraphics[width=1 \linewidth]{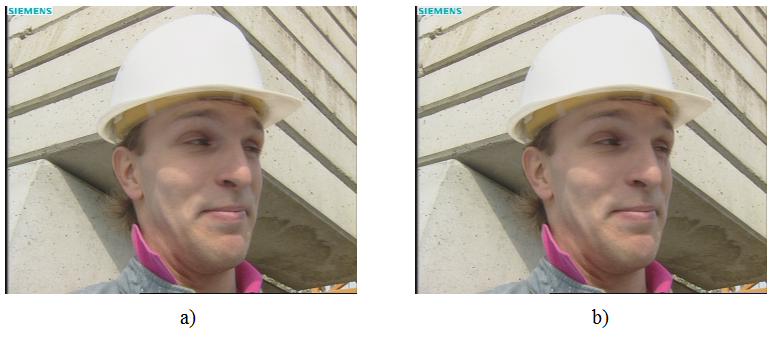}
						\caption{An example of a) original and b) watermarked frame.}
						\label{fig4}
					\end{figure}

					Our watermarking transparency is greater than the method proposed in \cite{park}. For better visualization, one original and watermarked frame of Foreman video sequence is shown in Figure \ref{fig4}. The measured PSNR for the frames which are shown in this figure is 47dB. This value of PSNR confirms the transparency of our watermarking method.
					
					\section{Conclusion}
					In this paper, we proposed a new robust watermarking method for uncompressed video data against compression attacks. We used the H.264/AVC and H.265/HEVC compression standards to evaluate the robustness of our method. To increase the robustness of watermarking against compression attack, we embedded the watermark into WBs. The WB is a block of a frame which is not similar to its corresponding blocks in the previous and next frames. Therefore, the coefficients of the corresponding residual block in the compression process would have high values and this way we could increase the robustness of the watermark. By simulation we measured the transparency and robustness of our proposed method and compared the results to the results of \cite{park}. We also verified the robustness of our proposed method by applying several compression attacks and each time the watermark could be extracted successfully with acceptable correlation with the original watermark. The results showed the proposed method has adequate transparency and robustness.


\begin{thebibliography}{1}
						\bibitem{bookw} {I. J. Cox, M. L. Miller, J. A. Bloom, J. Fridrich, and T. Kalker, \emph{Digital watermarking and steganography.} Morgan Kaufmann Publishers, Amsterdam, Boston, 2008.}
						
						\bibitem{zarmehi}{N. Zarmehi, M. Banagar, and M. A. Akhaee, ``Optimum decoder for an additive video watermarking with Laplacian noise in H.264,'' \emph{2013 10th International ISC Conference on Information Security and Cryptology (ISCISC)}, vol. 1, no. 5, pp. 29-30, August. 2013.}
						
						\bibitem{shih}{F. Y. Shih, \emph{Digital Watermarking and Steganography: Fundamentals and Techniques.} CRC Press, Inc., Boca Raton, FL, USA, 2007.}
						
						
						
						
						
						
						
						
						
						\bibitem{srv1}{X. Chang, W. Wang, J. Zhao, and L. Zhang, \emph{A survey of digital video watermarking.} Seventh International Conference on Natural Computation (ICNC), 2011.}
						
						\bibitem{srv2}{Jianjun Qin, \emph{The research of digital video watermarking.} Hunan Normal University 2010.}
						
						\bibitem{srv3}{N. Zarmehi and M. A. Akhaee, ``Video steganalysis of multiplicative spread spectrum steganography,'' \emph{22nd European Signal Processing Conference (EUSIPCO), Lisbon,} 2014, pp. 2440-2444.}
						
						\bibitem{srv4}{N. Zarmehi and M. A. Akhaee, ``Digital video steganalysis toward spread spectrum data hiding,'' \emph{IET Image Processing,} vol. 10, no. 1, pp. 1-8, 1 2016.}
						
						
						
						
						
						
						
						
						
						\bibitem{park}{H. Park, S. H. Lee, and Y. S. Moon, ``Adaptive video watermarking utilizing video characteristics in 3D-DCT domain,'' \emph{in Proceedings on the 5th International Workshop on Digital Watermarking, IWDW'06, Korea}, November 2006, vol. 4283 of Lecture Notes in Computer Science, pp. 397-406, Springer.}
						
						\bibitem{alavani}{M. A. Alavianmehr, M. Rezaei, M. S. Helfroush, and A. Tashk, ``A lossless data hiding scheme on video raw data robust against H.264/AVC compression,'' \emph{2nd International Conference on Computer and Knowledge Engineering (ICCKE), Mashhad}, pp. 194-198, October 2012.}
						
						
						\bibitem{masking}{P. Campisi and A. Neri, ``Video watermarking in the 3D-DWT domain
							using perceptual masking,'' in IEEE Int. Conf. Image Processing, 2005
							(ICIP 2005), pp. 997-1000.}
						
						\bibitem{obvious}{A. Koz and A. A. Alatan, ``Oblivious spatio-temporal watermarking
							of digital video by exploiting the human visual system,'' IEEE Trans.
							Circuits Syst. Video Technol., vol. 18, no. 3, pp. 326-337, Mar. 2008.}
						
						
						\bibitem{srv}{F. Hartung and B. Girod, ``Watermarking of uncompressed and compressed video,'' Signal Process., vol. 66, no. 3, pp. 283-301, May 1998.}
						
						
						
						\bibitem{h264}{ITU-T and ISO/IEC JTC 1, \emph{Advanced video coding for generic audiovisual services, ITU-T Rec. H.264 and ISO/IEC 14496-10 (AVC)}. version 1, 2003, version 2, 2004, versions 3, 4, 2005, versions 5, 6, 2006, versions 7, 8, 2007, versions 9, 10, 11, 2009, versions 12, 13, 2010, versions 14, 15, 2011, version 16, 2012.}
						
						\bibitem{h265}{B. Bross, W. -J. Han, G. J. Sullivan, J. -R. Ohm, and T. Wiegand, \emph{High Efficiency Video Coding (HEVC) text specification draft 9}, Document JCTVC-K1003, ITU-T/ISO/IEC Joint Collaborative Team on Video Coding (JCT-VC), October 2012.}
						
						\bibitem{rechardson1}{E. I. Richardson, \emph{The H.264 advanced video compression standard.} John Wiley \& Sons, Ltd, 2010.}
						
						\bibitem{rechardson2}{E. I. Richardson, \emph{H.264 and MPEG-4 video compression: Video coding for next-generation multimedia}, John Wiley \& Sons, Ltd, December 2003.}
					\end{thebibliography}
\end{document}